\documentclass[twocolumn,showpacs,preprintnumbers,amsmath,amssymb]{revtex4}

\usepackage{graphicx}
\usepackage{dcolumn}
\usepackage{bm}

\newcommand{\UG}{UGe$_2$}
\newcommand{\TX}{$T_X$}
\newcommand{\PX}{$P_X$}
\newcommand{\TS}{$T_{SC}$}
\newcommand{\PF}{$P_{FM}$}
\newcommand{\TF}{$T_{FM}$}

\newcommand{\et}{\textit{et al.}}

\begin{document}

\preprint{APS/123-QED}

\title{Stoner Gap in the Superconducting Ferromagnet \UG} 

\author{N. Aso,$^1$ G. Motoyama,$^2$ Y. Uwatoko,$^3$ S. Ban,$^4$ S. Nakamura,$^4$\\ 
T. Nishioka,$^5$ Y. Homma,$^6$ Y. Shiokawa,$^6$ K. Hirota,$^1$ and N.K. Sato$^4$}

\affiliation{%
$^1$Neutron Sci.~Lab., ISSP, University of Tokyo, Tokai, Ibaraki 319-1106, Japan\\
$^2$Department of Material Science, Graduate School of Material Science, University of Hyogo, Hyogo 678-1297 Japan\\
$^3$ISSP, University of Tokyo, Kashiwa 277-8581, Japan\\
$^4$Department of Physics, Graduate School of Science, Nagoya University, Nagoya 464-8602, Japan\\
$^5$Department of Material Science, Faculty of Science, Kochi University, Kochi 780-8520, Japan\\
$^6$Oarai Branch, Inst.~for Mater.~Research, Tohoku University, Oarai, Ibaraki 311-1313, Japan}

\date{\today}

\begin{abstract}
We report the temperature ($T$) dependence of ferromagnetic Bragg peak intensities and dc magnetization of the superconducting ferromagnet \UG~under pressure ($P$). We have found that the low-$T$ behavior of the uniform magnetization can be explained by a conventional Stoner model. A functional analysis of the data produces the following results: The ferromagnetic state below a critical pressure can be understood as the perfectly polarized state, in which heavy quasiparticles occupy only majority spin bands. A Stoner gap $\Delta(P)$ decreases monotonically with increasing pressure and increases linearly with magnetic field. We show that the present analysis based on the Stoner model is justified by a consistency check, i.e., comparison of density of states at the Fermi energy deduced from the analysis with observed electronic specific heat coeffieients. We also argue the influence of the ferromagnetism on the superconductivity.
\end{abstract}

\pacs{65.40.-b, 71.28.+d, 71.30.+h, 71.27.+a}
\maketitle

\section{Introduction}

Since a pioneer paper by Ginzburg on the coexistence of ferromagnetism and 
superconductivity \cite{Ginzburg57}, the interplay between these two long-range orderings 
has been an interesting topic in solid-state physics. Superconductivity and magnetism would be antagonistic because of the competitive nature between the superconducting screening (Meissner effect) and the internal fields generated by magnetic orderings. During the last three decades, however, the discovery of a number of magnetic superconductors 
has allowed for a better understanding of how magnetic order and superconductivity can coexist. 
It seems to be generally accepted that antiferromagnetism with local moments coming from rare-earth elements readily coexists with type II superconductivity. This is because superconductivity and magnetism are carried by different types of electrons; magnetism is connected with deeply seated 4\textit{f} electrons, while superconductivity is fundamentally related to the outermost electrons such as \textit{s}, \textit{p}, and \textit{d} electrons. 

In the case of a ferromagnetic superconductor, a trickier negotiation is needed 
for the coexistence, because internal fields are not canceled out in the range of a superconducting coherence length in contrast with an antiferromagnetic superconductor. In the classical magnetic superconductor ErRh$_4$B$_4$ with a superconducting transition temperature 8.7 K, for example, once the Er sublattice starts to order ferromagnetically below about 0.8 K, 
the superconductivity is immediately destroyed, except a very narrow coexistence region near 0.8 K \cite{Tachiki}. Here we note that the magnetic structure coexisting with the superconductivity is not purely ferromagnetic but spacially modulated. ErNi$_2$B$_2$C is a modern example of magnetic superconductor. A microscopic coexistence between weak ferromagnetism and superconductivity was reported, but detailed neutron diffraction investigations indicated that the magnetism coexisting with the superconductivity is not purely ferromagnetic \cite{Kawano02}, again. 
These examples seem to indicate that superconductivity hardly coexists with ferromagnetism, even though superconductivity and ferromagnetism are carried by different types of electrons. 
Recently, Saxena \et~discovered a new type of ferromagnetic superconductor \UG~in which superconductivity occurs at high pressures \cite{Saxena00}. It is particularly interesting to note that both of ferromagnetism and superconductivity may be carried by itinerant $5f$ electrons, which can be homogeneously spread in the real space, although it is still a matter of debate and remains to be resolved. This observation has renewed our interest on the interplay of ferromagnetism and superconductivity.

\begin{figure}[h]
\includegraphics[scale=0.65]{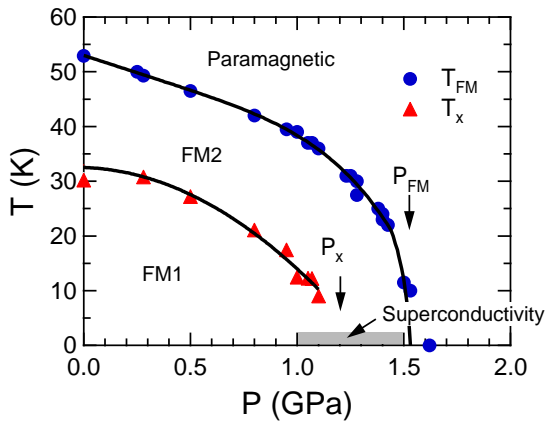}
\caption{\label{fig:diagram} 
(Color online) Phase diagram for \UG~determined by our neutron diffraction measurements. 
The shaded region between about 1.0 and 1.5 GPa shows a superconductivity region 
taken from the literature \cite{Nakane05}.
The solid lines are guides to the eye. ``FM1'' denotes a perfectly polarized ferromagnetic state 
in which only majority spin bands are occupied. For ``FM2'' state above \PX, see the discussion in the text.
}
\end{figure}

Figure \ref{fig:diagram} shows a temperature ($T$) vs pressure ($P$) phase diagram of \UG. A Curie temperature (\TF) is about 52 K at ambient pressure, and monotonically decreases with increasing pressure. Then it collapses to zero temperature at a ferromagnetic critical pressure \PF~($\sim$ 1.5 GPa). In the ferromagnetic phase, another phase transition or crossover seem to appear at \TX~($\simeq$ 32 K at ambient pressure). This characteristic temperature \TX~also decreases with increasing pressure and becomes suppressed to zero at a critical pressure \PX~($\sim$ 1.2 GPa). The transitions at \PX~and \PF~are likely of the first order in nature \cite{Pfleiderer02}. Superconductivity emerges in the pressure range between $\sim$ 1.0 and $\sim$ 1.5 GPa. Since a maximum superconducting transition temperature (\TS~$\sim$ 0.7 K) is observed at around \PX~\cite{Saxena00}, we speculate that the critical point \PX~plays an important role in the onset of the superconductivity (see, for example, Watanabe and Miyake \cite{Watanabe02}, Sandeman \et~\cite{Sandeman03}, and references therein). Very recently, Nakane \et~provided a supporting evidence for the speculation by means of ac magnetic susceptibility measurements under external magnetic fields $H$; in a plot of \TS~as functions of $P$ and $H$, the superconductivity always appears at around the critical point \PX, not around \PF~\cite{Nakane05}. However, there are still many unsolved questions in this unique material to be further clarified. To shed more light on the nature of the ferromagnetism as well as its pressure variation, we present in this paper the $T$-dependence of the uniform magnetization under pressure by the neutron diffraction technique, together with the dc magnetization method. 
Similar measurements were already reported, however the present experiment is so precise that we can analyze the functional dependence of the magnetization.
Actually, we have found that the low $T$-dependence of the uniform magnetization can be described by a conventional Stoner model. This enables us to extract new information about the ferromagnetism as follows: The low-$T$ and low-$P$ region of the ferromagnetic state, i.e., the FM1 region in Fig.~1, is understood as a perfectly polarized state in which only a majority spin band is occupied. As the pressure increases toward \PX, a Stoner gap $\Delta(P)$ in the heavy quasiparticle bands decreases monotonically, similarly to \TX$(P)$. When the pressure exceeds \PX, the gap seems to jump up, although the applicability of the Stoner model to this high pressure ferromagnetic state (FM2) is less convincing compared to the region $P <$ \PX. From these results, we argue the influence of an effective internal field produced by the ferromagnetism, which is found to be remarkably large below \PX, on the superconductivity.


\section{Experiment}

Single crystals were grown by the Czochralsky pulling method using a tetra-arc furnace installed 
at Oarai Branch of Institute for Material Research, Tohoku University \cite{Motoyama01}. 
The pressure was generated using a copper-beryllium (CuBe) based piston-cylinder clamp device \cite{Uwatoko} 
with Fluorinert FC-75 (3M Co. Ltd., Tokyo) as a pressure transmitting medium. 
The low temperature pressure was determined by measurements of the change 
in a lattice parameter of NaCl put together with the sample. Elastic neutron scattering experiments were done on the ISSP cold neutron triple-axis spectrometer HER (C1-1) installed at JRR-3M, JAERI, Japan, with a typical configuration of energy $k_{i}$ = 1.11 \AA $^{-1}$ or 1.555 \AA $^{-1}$ and collimations of Guide-Open-80'-80'. A cooled Be filter was placed 
before the sample to remove higher order contaminations. The crystals were oriented with the $a$-axis perpendicular to the scattering plane. Temperature was cooled down to 1.4 K using a $^4$He-pumping ILL-type orange cryostat. The dc magnetization measurements were carried out using a conventional vibrating sample magnetometer (VSM) \cite{Motoyama01}.

\section{Results and discussion}

\begin{figure}[h]
\includegraphics[scale=0.52]{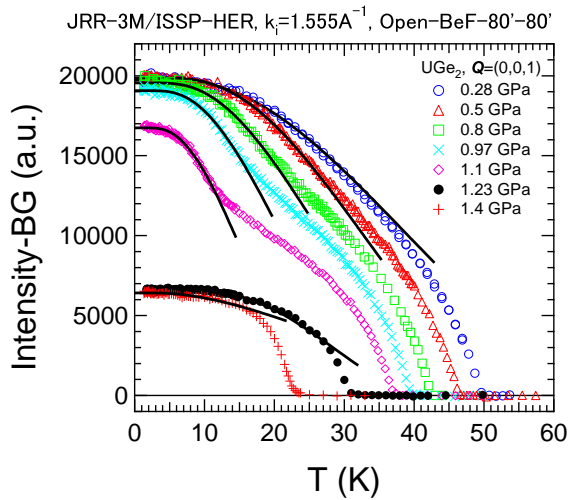}
\caption{\label{fig:Bragg} 
(Color online) Temperature dependence of the ferromagnetic Bragg peak intensities at $Q$ = (0,0,1) against the temperature $T$ 
measured at various pressures. 
``BG'' denotes background intensities in the paramagnetic phase of about 1250, which arise from 
the incoherent scattering of both the crystal itself and the pressure cell. Note that \PX~$\sim$ 1.2 GPa.
The solid lines are calculated results on the basis of the Stoner model described in the text. 
}
\end{figure}

In Fig.~\ref{fig:Bragg} we show the $T$-dependence of magnetic Bragg peak intensities $I_B(T)$ at $Q$ = (0,0,1) for several pressures. All data were accumulated at $k_{i}$ = 1.555 \AA $^{-1}$ in the process of increasing temperature. In contrast with a conventional dc magnetization measurement, neutron scattering experiments do not suffer from complications arising from a pressure cell contribution to the magnetization as well as a magnetic domain effect in a ferromagnetic sample, and hence the present results are not obscured at all by these effects. While there is no apparent anomaly in the curve of $P$ = 0.28 GPa, we clearly observe a steep increase below \TX~$\sim$ 10 K at 1.1 GPa. (In the present study, we define \TX~as a maximum temperature appearing in the second derivative of the $I_B(T)$ curve with respect to $T$; Note that this definition yields a \TX-value close to previously reported ones.) 
We note that the overall feature of the present result is consistent with the Bragg peak intensity and static magnetization data previously reported in Refs.~\cite{Huxley01, Tateiwa01b, Motoyama01}. At 1.23 GPa, such an anomalous behavior was not observed in accordance with \PX~$\sim$ 1.2 GPa.


Remembering that the neutron intensities are proportional to the square of magnetization $M$, 
we calculate the magnetic Bragg peak intensities in terms of the Stoner model, 
which is expressed as follows (see, for example, \cite{Stoner}); 
\begin{equation}
M = M_0 \{ 1 - \alpha \cdot T^{\frac{3}{2}} \cdot \exp(- \Delta / T ) \}, \label{eq:Stoner1}
\end{equation}
\begin{equation}
\alpha = \frac{3}{4} \sqrt{\pi} \{ \frac{1}{E_F} \} ^{\frac{3}{2}}, 
\Delta = 2 E_F \{ \frac{\Theta '}{E_F} -2^{-\frac{1}{3}} \}, \label{eq:Stoner2}
\end{equation}
where $M_0$ indicates the magnetization at zero temperature, $\Delta $ a so-called Stoner gap, 
$E_F$ a Fermi energy, and $\Theta '$ is a molecular field coefficient. 
The results are shown in Fig.~\ref{fig:Bragg} by solid lines.
Interestingly, we find good agreement between the low-$T$ magnetization data and the calculation. 
(The observation of the exponential like $T$-dependence of the magnetization, instead of a conventional $T$-power law behavior due to spin wave excitaions, is probably related to a huge uniaxial magnetic anisotropy of \UG.) This agreement suggests that the decrease in the magnetization at low temperatures is mainly caused by electron-hole excitations in quasiparticle bands. 


\begin{figure}[htbp]
\includegraphics[scale=0.55]{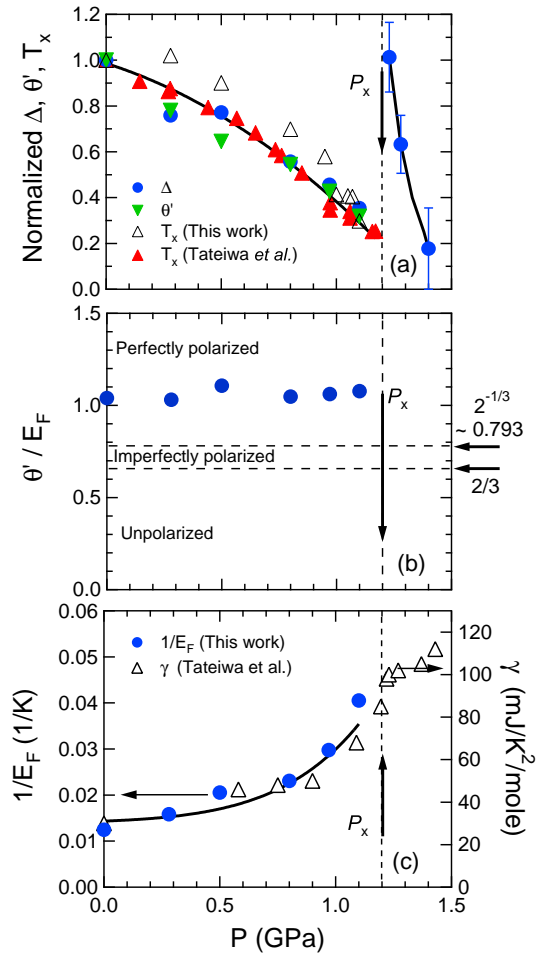}
\caption{\label{fig:parame} 
(Color online) Pressure dependence of the obtained parameters from the Stoner model. 
The solid lines are guides to the eye. (a) $\Delta $, $\Theta ' $, and \TX~plotted are 
normalized with respect to a respective ambient pressure value; 
$\Delta $ = 39.5 K, $\Theta ' $ = 83.4 K, and \TX~= 30.2 K. 
We also plot \TX~taken from Ref.~\cite{Tateiwa01a}. 
(b) Ratio of $\Theta '$/$E_F$ is plotted against $P$ below \PX. 
Note that the pressure region of $P$ $<$ \PX~corresponds to the perfectly polarized state 
in the Stoner model, i.e., $\Theta '$/$E_F$ $>$ 2$^{-1/3}$. 
For the region above \PX, see the discussion in the text.
(c) Inverse Fermi energy 1/$E_F$ is plotted against $P$ below \PX, together with an electronic specific heat coefficient $\gamma $ taken from Ref.~\cite{Tateiwa01a}.
}
\end{figure}


From the least square fitting of the data, we estimate a set of parameters $\alpha $ and $\Delta $ in eq.~(\ref{eq:Stoner1}), which further enables us to evaluate $E_F$ and $\Theta '$ using eq.~(\ref{eq:Stoner2}). First we concentrate on the pressure region below \PX. In Fig.~\ref{fig:parame} (a) we show $\Delta $ and $\Theta '$ together with \TX, each of which is normalized to unity at ambient pressure. It is interesting to note that these quantities seem to lie on a single line, suggesting that the characteristic energy scale of unknown origin, \TX, is related to the Stoner gap $\Delta $ (equivalently $\Theta '$).

In Fig.~\ref{fig:parame} (b) we plot a ratio of $\Theta '$/$E_F$ against $P$. According to the Stoner model, the ratio greater than 2$^{-\frac{1}{3}} (\sim $ 0.793) means that the system is in a perfectly polarized ferromagnetic state, where only a majority spin band is occupied. When the ratio lies between 2/3 and 2$^{-\frac{1}{3}}$, an imperfectly polarized ferromagnetic state occurs,  where a minority spin band becomes to be partially occupied by quasiparticles. Further in the case that the ratio is smaller than 2/3, the system is paramagnetic. As seen in the figure, our analysis indicates that the perfectly polarized state is realized below \PX. This result is supported by band structure calculations indicating that Fermi surfaces have a predominantly majority spin character \cite{Yamagami03, Picket}.

In Fig.~\ref{fig:parame} (c) we plot an inverse of Fermi energy 1/$E_F$ deduced from the above analysis as a function of $P$. To estimate a density of states at the Fermi energy, $D$($E_F$), we assume that $E_FD$($E_F$) is a constant value independent of pressure. Then, 1/$E_F$ corresponds to $D$($E_F$). On the other hand, $D$($E_F$) can be directly obtained from an electronic specific heat coefficient $\gamma $, which is also shown in Fig.~\ref{fig:parame} (c) for camparison \cite{Tateiwa01a}.
As is clearly seen, 1/$E_F$ is proportional to $\gamma $, i.e., $1/E_F = c \gamma $ with a $P$-independent constant $c$. This coincidence justifies our interpretation based on the Stoner model.

\begin{figure}[t]
\includegraphics[scale=0.38]{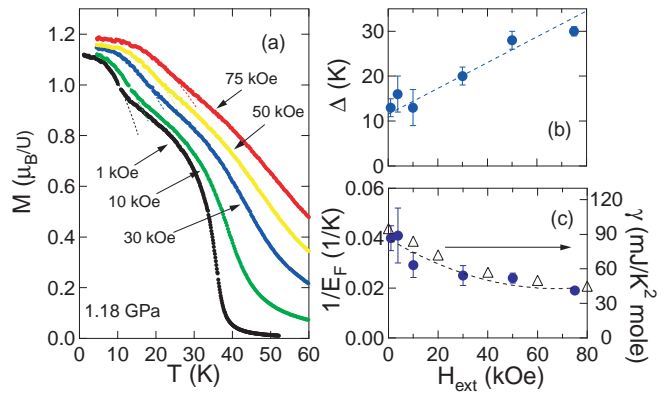}
\caption{\label{fig:field} 
(Color online) (a) Temperature dependence of the static magnetization at 1.18 GPa under various magnetic
fields along the magnetization easy $a$-axis. (b) Stoner gap $\Delta$ is plotted as a function of 
the external magnetic field $H_{ext}$. 
(c) Inverse Fermi energy 1/$E_F$ is plotted together with the electronic specific heat coefficient $\gamma $ 
measured at 1.15 GPa taken from Ref.~\cite{Tateiwa02}. Note that 1/$E_F$, which is proportional to 
the density of states at $E_F$, agrees well with the observed $\gamma$ value.
}
\end{figure}

Figure \ref{fig:field} (a) shows the dc magnetization $M(T)$ at 1.18 GPa ($<$ \PX) under external magnetic fields $H_{ext}$. (The magnetization of the pressure cell was subtracted from the total measured magnetization.) The magnetic field was applied along the magnetization easy $a$-axis. We observed that the $M(T)$ curve shows a step-like increase at lower fields similarly to the $I_B(T)$ curve, and that \TX~exhibits an increase with $H_{ext}$ in accordance with the previous results \cite{Pfleiderer02,Nakane05}. We find that the static low-$T$ magnetization can also be well described by the Stoner model (see dotted lines). 

In Fig.~\ref{fig:field} (b) we plot $\Delta $ as a function of $H_{ext}$, which was obtained in the same manner as above. It is found that $\Delta $ increases almost linearly with $H_{ext}$, as shown by a broken line. This is highly expected from the Stoner model; the gap in the quasiparticle bands should linearly increase with the magnetic field due to the Zeeman effect as follows, 

\begin{equation}
\Delta (H) = \Delta + 2 g \mu _B S H.
\end{equation}
Here, $\Delta $ is a value at zero magnetic field, i.e., $\Delta (H = 0)$, 
$g$ and $S$ denote a $g$-factor and the magnitude of the quasiparticle spin, respectively, and
$\mu_B$ is the Bohr magneton. Indeed, the value of $\Delta $ $\simeq$ 12 K estimated from the extrapolation to zero field is consistent with a value obtained from the Bragg peak intensity (at $H$ = 0) mentioned above.
A set of parameters, $g$ = 6/7 and $S$ = 5/2 corresponding to an \textit{f} electron, 
produces better agreement between the observation and the calculation than a different set of parameters, $g$ = 2 and $S$ = 1/2 
for a free electron. This may reflect that the heavy quasiparticle arises from \textit{f} electrons. 

The slope of the broken line in Fig.~4 (b) is calculated to be about 0.3 K/kOe. It is very interesting to note that this value is almost the same as the slope of curves in a plot of \TX~vs $H$ (see, for example, Ref.\cite{Watanabe02} and references therein). This clearly supports that \TX~is related to the Stoner gap $\Delta$, as mentioned above.


Figure~\ref{fig:field} (c) shows the $H_{ext}$-dependence of $1/E_F$ (at 1.18 GPa) obtained from the least square fitting of the $M(T)$ data to the Stoner model. We also plot reported values of the $\gamma $-coefficient observed at 1.15 GPa under external magnetic fields \cite{Tateiwa02}. Again, we find the same relation $1/E_F = c \gamma $ with the same scale factor $c$ as the above. Note that there is no adjustable parameter at all, nevertheless we find the good agreement between $D(E_F)$ estimated from the Stoner model and deduced from the heat capacity experiments. This clearly proves the validity of our model analysis.

Let us return to the pressure region of $P >$ \PX. As may be seen from Fig.~\ref{fig:Bragg}, the accordance between the calculated and the experimental results is less convincing compared with that for $P <$ \PX, for which there are two possible explanations: First, the low $T$-dependence of the uniform magnetization (for $P >$ \PX) can no longer be described by the Stoner model. Second, the Stoner model is still applicable to the FM2 region, but a pressure distribution within the sample will cause the $M(T)$ curve to deviate from the Stoner model. (The Curie temperature decreases steeply above \PX~(see Fig.~1). In such a case, the experimetal results could be obscured by even a small pressure distribution within the sample~\cite{Uwatoko,Sidorov05}.) 
Since it is unclear which is dominant, we tentatively tried to apply the Stoner model to the FM2 region. The obtained $\Delta$-values are as follows; $\Delta$ = 40 ($\pm $ 6), 25 ($\pm $ 5), and 7 ($\pm $ 7) K at $P$ = 1.23, 1.28, and 1.40 GPa, respectively. We find that $\Delta$ shows a jump near at \PX~on going from the FM1 to FM2 region, and that $\Delta$ decreases monotonically with further increasing pressure and finally tends toward zero in the vicinity of the ferromagnetic critical pressure \PF. Note that the jump of $\Delta$ reflects the sudden change in the solpe of the $I_B(T)$ curves below and above \PX.
A further investigation is needed to clarify whether the magnetization in the FM2 region can be described by the Stoner model or not.

Finally we discuss the correlation between the ferromagnetism in the FM1 region and the superconductivity.
It is evident that the Stoner gap behavior in the FM1 region is firmly established by the consistency check, i.e., the comparison of our results with the $\gamma$-coefficients. Using the parameter $\Theta '$ in eq.~(2) obtained from the fitting, we can estimate an effective internal field $H_{eff}$ seen by the itinerant electrons due to the ferromagnetism by the definition of $\mu_BH_{eff}=k_B\Theta '$.
Then we find it to be very large; for example, $H_{eff}$ $\ge$ 100 T at ambient pressure and $H_{eff}$ $\sim$ 40 T at 1.1 GPa (see Ref.~\cite{Sato05} for detail). This may explain an asymmetric shape of the superconducting dome with respect to \PX~in the $T$-$P$ phase diagram, if we assume that the superconductivity does not survive under such a strong internal field. In the literature, it has been speculated that the nonunitary superconducting state would be realized in \UG; otherwise the superconductivity would not coexist with the feromagnetism. However, it seems very unlikely that the strong internal field mentioned above dose not kill the superconductivity, even if the spin-triplet pairing state would be formed. 
This leads us to suggest spatially inhomogeneous coexistence of the ferromagnetism and the superconductivity, provided that the superconductivity below \PX~is intrinsic, but not due to the pressure inhomogeneity. We need a further experiment to confirm this possibility.

\section{summary}

We investigated the uniform magnetization of the pressure-induced superconductor \UG~by the neutron diffraction technique together with the dc magnetization measurements under pressure. For this strongly anisotropic ferromagnet, we found that the low-$T$ behavior in the magnetization of the FM1 region can be explained by the conventional Stoner model. Our analysis based on the Stoner model produces the following results: The ferromagnetic state below the critical pressure \PX~($\sim$ 1.2 GPa) is understood as the perfectly polarized state in which the heavy quasiparticles occupy only the majority spin band. The Stoner gap $\Delta$ in the heavy quasiparticle bands was estimated to be about 40 K at ambient pressure, and $\Delta$ was found to decrease monotonically with increasing pressure $P$ and to increase linearly with magnetic field $H$. The similarity between the $P$- and $H$-dependences of $\Delta$ and \TX~suggests that the characteristic energy \TX~of unknown origin can be related to the Stoner gap.
Assuming that the product $E_FD$($E_F$) is constant, we evaluated the $P$- and $H$-dependence of a density of states at the Fermi energy $D$($E_F$) using the Stoner model. Then we found that $D$($E_F$) = $c$$\gamma$ with the same constant $c$ for both the $P$- and $H$-dependence of the electronic specific heat coefficient $\gamma $. This justifies our interpretation based on the Stoner model. Finally we argued  the relationship between ferromagnetism and superconductivity; the effective internal field seen by itinerant electrons is estimated to be sufficiently strong that the superconductivity would hardly survive, which leads us to suggest the spatially inhomogeneous coexistence of ferromagnetism and superconductivity. We hope that these results stimulate further theoretical investigations.

\begin{acknowledgments}
We thank K.~Miyake and S.~Watanabe for useful discussions. 
This work was supported by a Grant-in-Aid from the Ministry of 
Education, Culture, Sports, Science and Technology, Japan. 
NKS also thanks Daiko Foundation for partial financial support.
\end{acknowledgments}


\end{document}